\documentclass[prd,11pt,english,nofootinbib,amsfonts,amssymb,superscriptaddress]{revtex4} 
\usepackage{amsmath}
\usepackage{babel}
\usepackage{bm}
\usepackage{enumerate}

\begin{document}
\newcommand\new[1]{\ensuremath{\blacktriangleright}#1\ensuremath{\blacktriangleleft}}
\newcommand\note[1]{$\blacktriangleright$[\emph{#1}]$\blacktriangleleft$}
\newcommand\Det{\textrm{Det}}

%%%%%%%%%%%%%%%%%%%%%%%%%%%%%%%%%%%%%%%%%%%%%%%%%%%%%%%%%%%%%%%%%%%%%%%%%%%%%
\title{Brans--Dicke geometry}

\author{Raffaele Punzi}
\email{raffaele.punzi@desy.de}
\affiliation{Zentrum f\"ur Mathematische Physik und II. Institut f\"ur Theoretische Physik, Universit\"at Hamburg, Luruper Chaussee 149, 22761 Hamburg, Germany}

\author{Frederic P. Schuller}
\email{fps@aei.mpg.de}
\affiliation{Max Planck Institut f\"ur Gravitationsphysik, Albert Einstein Institut, Am M\"uhlenberg 1, 14467 Potsdam, Germany}

\author{Mattias N.\,R. Wohlfarth}
\email{mattias.wohlfarth@desy.de}
\affiliation{Zentrum f\"ur Mathematische Physik und II. Institut f\"ur Theoretische Physik, Universit\"at Hamburg, Luruper Chaussee 149, 22761 Hamburg, Germany}

\begin{abstract}
We reveal the non-metric geometry underlying $\omega\rightarrow 0$ Brans-Dicke theory by unifying the metric and scalar field into a single geometric structure. Taking this structure seriously as the geometry to which matter universally couples, we show that the theory is fully consistent with solar system tests. This is in striking constrast with the standard metric coupling, which grossly violates post-Newtonian experimental constraints.
\end{abstract}
\maketitle

%%%%%%%%%%%%%%%%%%%%%%%%%%%%%%%%%%%%%%%%%%%%%%%%%%%%%%%%%%%%%%%%%%%%%%%%%%%%%%
Brans-Dicke gravity theory aims at describing the dynamics of a spacetime metric $g$ by employing an additional scalar degree of freedom $\phi$ in order to model a dynamical gravitational constant~\cite{Brans:1961sx}. Brans-Dicke theory and, more generally, scalar tensor theories of gravity, have many interesting properties, and have been extensively discussed in the literature. Perhaps the most fruitful area of their application is cosmology, e.g. in~\cite{Carroll:2004hc,Sen:2000zk,Bertolami:1999dp,Holden:1999hm,Torres:2002pe}, where the scalar field is often employed as a quintessence field to drive accelerating phases of the universe; scalar-tensor theories naturally appear in brane-world scenarios~\cite{Garriga:1999yh,Brax:2004xh}, or arise as equivalent formulations of $f(R)$ gravity theories with Ricci scalar corrections~\cite{Chiba:2003ir,Hindawi:1995cu,Magnano:1987zz,Deruelle:2008fs}.

The original family of Brans-Dicke actions is
\begin{equation}
  S_\omega[g,\phi] = \int d^4x \sqrt{-g}\Big[\phi R - \omega \phi^{-1} g^{-1}(d\phi,d\phi)\Big],
\end{equation}
parameterized by the dimensionless parameter $\omega$. This is completed into a theory of gravity by the prescription that matter couple to the metric $g$ only, but not to the scalar field $\phi$. While the theory as such is not inconsistent or experimentally falsified, the long history of its study has turned up a number of concerns, that lessen the appeal of Brans-Dicke theory, and more general scalar tensor theories, as alternatives to general relativity:

\vspace{3pt}\noindent{\it Problem of naturalness:} there is no fundamental principle that distinguishes the form of the Brans-Dicke action, or indeed any other scalar-tensor theory. In constrast, general relativity is distinguished as the unique metric gravity theory with second order field equations, up to a cosmological constant.

\vspace{3pt}\noindent{\it Problem of indeterminacy:} there are no principles dictating the value of the Brans-Dicke parameter~$\omega$, nor experimental results bounding it away from the Einstein limit $\omega\rightarrow\infty$. In contrast, the only free parameter in Einstein gravity, the cosmological constant, is nowadays very precisely bounded from both sides.

\vspace{3pt}\noindent{\it Problem of experimental consistency:} increasing precision of solar system tests alone have shifted~$\omega$ over the years by many orders of magnitude to now over $4\cdot 10^4$ \cite{WillReview}. In contrast, the predictions of general relativity have remained consistent with increasingly precise experimental data in the solar system over the decades. Also in more general scalar tensor theories the additional scalar fields usually turn out to be very dangerous for the consistency of the gravity theory with solar system observations \cite{Erickcek:2006vf}.

\vspace{3pt}\noindent{\it Problem of geometric interpretation:} no geometric meaning is attached to the pair of background fields $(g,\phi)$, which could explain the specific interplay of the metric and the scalar field in the gravitational part of the action and justify a particular coupling prescription for matter. In contrast, the understanding of the gravitational degrees of freedom in general relativity as the components of a single metric tensor allows for a compelling geometric formulation of the theory.

\vspace{3pt}In this letter, we show that all of the above problems are related, and indeed can be resolved in one stroke, by combining the metric and scalar field into a gravitational multiplet in form of a higher rank geometric structure. From this fact everything else follows without further assumptions. In particular, we will demonstrate that the refinement of Einstein-Hilbert gravity based on this higher rank tensor naturally singles out $\omega\rightarrow 0$ Brans-Dicke theory under all scalar-tensor theories, but also requires a specific coupling of matter to the data $(g,\phi)$, which is different from the standard coupling. The central point of this letter is that, in striking contrast with the standard coupling to matter, the theory now agrees precisely with general relativity in the solar system, up to the experimentally accessible first post-Newtonian order.

We now make the above technically precise. The pivotal geometric construction is the definition of the fourth rank tensor field
\begin{equation}\label{almostmetric}
G^{abcd} = g^{ac}g^{bd} - g^{ad}g^{bc} + \tilde\phi (-g)^{-1/2} \epsilon^{abcd}\,,
\end{equation}
where $\epsilon$ is the Levi-Civita tensor density with $\epsilon^{0123}=-1$, and $\tilde\phi$ is a function of $\phi$, whose precise form (\ref{tphi}) will be determined shortly.
This fourth rank tensor encodes the scalar-tensor data $(g,\phi)$ in a geometrically distinguished way:
the tensor field $G^{abcd}$ has an inverse $G_{abcd}$ in the sense that locally $G^{abmn} G_{mncd} = 4\delta^{[a}_c\delta^{b]}_d$, and both~$G$ and its inverse share the symmetries ${G_{abcd}=G_{cdab}}$ and $G_{abcd}=G_{[ab][cd]}$. These properties identify (\ref{almostmetric}) as a special case of an (inverse) area metric on the manifold~$M$, see \cite{Schuller:2005ru}. Indeed, $G_{abcd}X^aY^bX^cY^d$ yields the area squared of a parallelogram spanned by vectors $X$ and $Y$ at the same point, as measured by the metric $g$, wherever on $M$ the scalar field $\tilde\phi$ vanishes. Conversely, a non-zero value for $\tilde\phi$ modifies the area measure at a point in a way that could not be achieved by a different metric alone, since that could not affect the totally antisymmetric part of $G$.

Area metric differential geometry now gives us excellent technical control over this structure. Employing, in particular, the recent construction of an area metric volume form $\omega_G$ and curvature scalar $R_G$, one immediately finds the area metric refinement of the Einstein-Hilbert action,
\begin{equation}\label{GEH}
   S[G] = (2\kappa)^{-1} \int \omega_G \, R_G\,,
\end{equation}
whose formulation obviously does not require the introduction of any new parameters. Variation of this action with respect to a generic area metric $G$ yields equations of motion, which for the area metrics (\ref{almostmetric}) of interest to this paper reduce to the vacuum field equations of Brans-Dicke theory for $\omega\rightarrow 0$, identifying
\begin{equation}\label{tphi}
\phi = (2\kappa)^{-1} (1+\tilde\phi^2)^{-1/2}\,.
\end{equation}
 For full technical detail of the area metric variation of the Einstein-Hilbert action, we refer the reader to \cite{Punzi:2006nx}. Thus at the level of vacuum field equations, Brans-Dicke theory with vanishing parameter~$\omega$ is singled out as the area metric refinement of Einstein-Hilbert theory for an area metric defined by (\ref{almostmetric}).

We emphasize that without specifying the coupling of matter to the gravitational degrees of freedom, any dynamics for the latter are void of physical meaning; not even vacuum solutions can be interpreted without studying the motion of matter~\cite{Magnano:1993bd}. Indeed, it is the question of the matter coupling which truly distinguishes the otherwise equivalent views of $\omega\rightarrow 0$ Brans-Dicke theory as dynamics for a metric or an area metric spacetime. We will show that coupling matter minimally to the area metric multiplet renders the theory consistent with classical tests. To this end, but also for further theoretical considerations, we now explore the subtle issue of matter coupling in some detail.

 Taking seriously the intriguing role the area metric point of view plays in the vacuum theory, we include a matter action $S_m[G,\Psi]$ for matter fields~$\Psi$. By  variation with respect to $G$ we obtain field equations of the form $K_{abcd} = T_{abcd}$, where the gravitational tensor $K$ and the source tensor $T$ are the functional derivatives of the gravity action $S$ and the chosen matter action $S_m$, respectively. With the Brans-Dicke ansatz (\ref{almostmetric}) for the area metric, the tensor $K$ reduces algebraically to a scalar and a second rank tensor. For matter with source tensor
\begin{equation}\label{simsource}
T_{abcd}=2 T_{[a[c}g_{d]b]}-\frac{1}{3}Tg_{a[c}g_{d]b}-\frac{1}{24}\bar T\sqrt{-g}\epsilon_{abcd}\,,
\end{equation}
where $T=g^{ab}T_{ab}$ and $\epsilon_{0123}=1$, the fourth rank equation reduces to a pair of equations, one scalar and one second rank tensor equation \cite{Punzi:2006nx,Schuller:2007ix}:
\begin{eqnarray}
G_{ab} & = & \frac{1}{\phi}\left(\nabla_a\partial_b\phi-g_{ab}\square \phi\right) + \kappa \Big(4 T_{ab}+\frac{1}{2}\tilde\phi g_{ab}\bar T\Big),\nonumber\\
3\square \phi & = & 4\phi\kappa T + \frac{1-8\kappa^2\phi^2}{2\left(1-4\kappa^2\phi^2\right)^{1/2}}\bar T\,.\label{BDeqs}
\end{eqnarray}
Note that while the standard $\omega\rightarrow 0$ Brans-Dicke equations are recovered in vacuo, the matter coupling is non-standard, but
precisely of the form required by our identification of the area metric as the gravitational degrees of freedom, and the thus induced definition of the source tensor as the functional derivative of the matter action with respect to the area metric.
This structurally coherent inclusion of matter completes the specification of all elements of the theory at a formal level, and we now turn to physical predictions.

Applications to the geometrically simplest case of a spatially homogeneous and isotropic background, and the relevance of the emerging refined notion of cosmological perfect fluids, described by three rather than two macroscopic parameters, have been discussed in earlier work~\cite{Schuller:2007ix,Punzi:2006hy}. Here, we will address the crucial question of the compatibility of the theory with solar system experiments, which in general is a delicate issue in theories with additional scalar fields~\cite{Erickcek:2006vf}.

We will demonstrate that the area metric interpretation of the Brans-Dicke data ensures precise agreement with general relativity to first post-Newtonian order, and thus passes the classical solar system tests. In order to see this, we employ the result that the local null structure of area metric manifolds~\cite{Hehlbook} is governed by the totally symmetric Fresnel tensor
\begin{equation} 
\mathcal{G}_{G\,abcd} = -\frac{1}{24}\omega_{\hat G}^{ijkl}\omega_{\hat G}^{mnpq}\hat G_{ijm(a}\hat G_{b|kn|c}\hat G_{d)lpq}\,,
\end{equation}
which is fully determined by the cyclic part $\hat G_{abcd}=G_{abcd}-G_{[abcd]}$ of the area metric $G$. Moreover, the propagation of light in the geometric-optical limit of Maxwell theory on an area metric background is governed by stationary paths $x$ of the functional
\begin{equation}\label{Finsler}
   L[x] = \int d\tau\, \mathcal{G}_G(\dot x, \dot x, \dot x, \dot x),
\end{equation}
that are also $\mathcal{G}_G$-null, as was shown from first principles in~\cite{Punzi:2007di}. In the point particle idealization, planetary motion is described by non-null geodesics in the same Finsler geometry defined by~$L[x]$, see~\cite{Punzi:2007di}. For our Brans-Dicke geometry (\ref{almostmetric}), one finds that the Fresnel tensor takes the simple form
\begin{equation}
  \mathcal{G}_G(\dot x, \dot x, \dot x, \dot x) = \left(2\kappa\phi\,g(\dot x, \dot x)\right)^2.
\end{equation}
This implies that the Finsler geodesics derived from~(\ref{Finsler}) coincide with the geodesics of the conformally rescaled metric
\begin{equation}\label{testmetric}
 g_{\textrm{test}} = 2\kappa\phi\, g\,,
\end{equation}
which is thus the effective background seen by light and massive test particles. This fact immediately allows us to apply the post-Newtonian formalism for a comparison of the predictions of the theory with those of general relativity.

We define post-Newtonian parameters as usual by an expansion of the metric seen by light and massive test particles in terms of the Newtonian potential $U$,
 \begin{eqnarray}\label{postNewt}
g_{\textrm{test}} & = &  {}-(1+2U+2\beta U^2)dt^2\nonumber\\
& & {}+ (1-2\gamma U)(dr^2+r^2 d\Omega^2)\,,
\end{eqnarray}
assuming a spherically symmetric situation. The parameters $\beta$ and $\gamma$ displayed here are the relevant parameters for testing theories without preferred-frame effects, with global conservation of momentum, in the solar system range. General relativity corresponds to $\beta=\gamma=1$; any departure from these values is tightly constrained. The best current bound for $\gamma$ comes from Doppler tracking of Cassini, and is $|\gamma-1|<2.3\cdot 10^{-5}$, while data on the perihelion shift of Mercury yields the bound~$|\beta-1|<3\cdot 10^{-3}$~\cite{WillReview}.

The post-Newtonian parameters for our theory are now easily obtained from the well-known static spherically symmetric vacuum solutions of $\omega\rightarrow 0$ Brans-Dicke theory \cite{Brans:1961sx,Brans:1962}, which take the form
\begin{equation}
g = -e^{2\alpha(r)}dt^2 + e^{2\beta(r)}(dr^2+r^2 d\Omega^2)
\end{equation}
in isotropic coordinates. The functions $\alpha,\beta$ and the Brans-Dicke scalar $\phi$ depend on $r$ as
\begin{eqnarray}
&& e^\alpha(r)=e^{\alpha_0}f(r)^\lambda\,,\quad e^\beta(r)=e^{\beta_0}h(r)^2f(r)^{1-\lambda(1+C)}\,,\nonumber\\
&& \phi(r)=\phi_0 f(r)^{\lambda C}
\end{eqnarray}
in terms of the functions
\begin{equation}
f(r)=\frac{1-B/r}{1+B/r}\,,\quad h(r)=1+B/r\,,
\end{equation}
and constants $\alpha_0,\beta_0,\phi_0,B,C$, and we used the shorthand ${\lambda=(C^2+C+1)^{-1/2}}$.
Requiring that the effective metric $g_{\textrm{test}}$ reduces to the Minkowski metric at spatial infinity implies that ${e^{-2\alpha_0}=e^{-2\beta_0}=2\kappa\phi_0}$. The expansion of $g_{\textrm{test}}$ in powers of $B/r$ and comparison with (\ref{postNewt})
yields the Newtonian potential $U(r)=-M/r$ with central mass $M=\lambda(C+2)B$, and
\begin{equation}
\beta=1\,,\quad \gamma=1\,.
\end{equation}
This is in precise agreement with general relativity at first post-Newtonian order, so that solar system tests are passed with flying colors by the entire family of vacuum Brans-Dicke solutions, independent of value of the integration constant $C$. This is in pleasant contrast to the problems with the commonly stipulated coupling of matter to the metric data only, which gives~$\beta=1$, but ${\gamma=(\omega+1)/(\omega+2)}$, and is utterly inconsistent with $\omega\rightarrow 0$ Brans-Dicke dynamics. The fact that conformal changes in the matter coupling, precisely of the form (\ref{testmetric}), may restore observational consistency in scalar tensor theories has not gone unnoticed in the literature, see e.g.~\cite{Deruelle:2008fs,Magnano:1993bd}. Indeed, it can be verified that the theory studied here appears from this viewpoint as Einstein gravity for a metric $g_\textrm{test}$, with a scalar field $\phi$ and point-like matter both minimally coupled to~$g_\textrm{test}$. The key point here is that the area metric structure is recognized as a geometric principle which distinguishes this matter coupling (and predicts a different coupling to non-pointlike matter, such as gauge fields), and ensures that the $\omega\rightarrow 0$ theory is as consistent with observational data in the solar system as general relativity.

For completeness, we remark that the interior solution for any static spherically symmetric source may be matched to precisely one member of the above family of vacuum solutions. Consider, for instance, a weakly self-gravitating body, modelled by a non-interacting fluid described by its energy density only. Such fluids in area metric backgrounds were studied in \cite{Punzi:2007di}, and found to be composed of idealized point particles moving along the non-null Finsler geodesics discussed above.
Using such a source, the gravity equations (\ref{BDeqs}) simplify to
\begin{equation}
G_{ab} = \frac{1}{\phi}\nabla_a\partial_b\phi + \frac{16\kappa^2}{3}\tilde\rho \phi u_au_b\,,\quad\square\phi = 0\,,
\end{equation}
where $\tilde \rho$ is the energy density parameter of the fluid and~$u$ is its velocity field. We now match, at the boundary~${r=R}$ of the source, the integration constants of the exterior solution to integrals over appropriate components of the energy of the source. This can be done analytically in the weak field approximation. Thus we find the relations $C=0$, $\lambda=1$ and the central mass $M=2B$ as
\begin{equation}
M=\int_0^Rdx\,x^2 \frac{2\tilde\rho(x)}{3\phi_0}\,.
\end{equation}
The thus defined exterior solution is precisely the Schwarzschild solution in isotropic coordinates; apart from conventional factors, the identification of the mass is standard. This exemplary calculation easily generalizes for any static spherically symmetric source, not necessarily leading to the Schwarzschild solution, but with all integration constants determined by integrals over energy-momentum tensor components of the respective source. Thus matching exterior vacuum solutions to interior solutions for matter admitted by the Brans-Dicke geometry~(\ref{almostmetric}) is always possible, and the motion of test particles is in agreement with general relativity up to at least first post-Newtonian order.

\vspace{3pt}{\it Conclusion.} The area metric perspective adopted in this Letter successfully resolves a number of pertinent questions in the context of Brans-Dicke and more general scalar-tensor theories. Brans-Dicke gravity with vanishing Brans-Dicke parameter $\omega\rightarrow 0$ is singled out among all scalar tensor theories of gravity as the simplest area metric refinement of Einstein-Hilbert gravity. As such it is a rigid extension of Einstein-Hilbert gravity without additional freely adjustable parameters in the action. At the level of the vacuum equations this observation amounts to little more than a mathematical peculiarity, but this new geometric view of the theory leads to profound physical consequences: regarding the area metric multiplet~(\ref{almostmetric}) as the gravitational degrees of freedom, rather than the metric~$g$ and the scalar field~$\phi$ individually, requires that matter couple directly to the area metric. The refined geometric background then results in a refined notion of perfect fluids, as needed, for example, in the context of cosmology and planetary models in the solar system. The dynamics of standard model matter for which area metric spacetimes provide an equally good habitat are subtly generalized: for instance, the coupling of gauge theories to area metric backgrounds implies that light rays follow geodesics in a Finsler geometry induced by the area metric. It is the interplay of the gravitational dynamics and the matter coupling to the Brans-Dicke geometry, which makes the resulting $\omega\rightarrow 0$ theory fully consistent with all solar system tests.

The success of the area metric interpretation of Brans-Dicke theory may be taken as a hint towards a more fundamental relevance of area metric spacetimes. From this point of view, models of the solar system might arise from sources more complicated than~(\ref{simsource}), which would yield area metric backgrounds that cannot be written in the the simple Brans-Dicke form~(\ref{almostmetric}). This raises the issue of possible observable effects; one is tempted to speculate whether a full area metric treatment could even explain effects such as dark matter or the Pioneer anomaly in some equally natural fashion.

%%%%%%%%%%%%%%%%%%%%%%%%%%%%%%%%%%%%%%%%%%%%%%%%%%%%%%%%%%%%%%%%%%%%%%%%%%%%%%
\acknowledgments
The authors wish to thank Nathalie Deruelle, J\"org Kulbartz and Marcelo Salgado for helpful discussions and comments. RP and MNRW acknowledge full financial support from the German Research Foundation DFG through MNRW's Emmy Noether fellowship grant WO 1447/1-1.

%%%%%%%%%%%%%%%%%%%%%%%%%%%%%%%%%%%%%%%%%%%%%%%%%%%%%%%%%%%%%%%%%%%%%%%%%%%%%%

\end{document}